\documentclass[amsmath,amssymb,amsbsy,reprint,prb,preprintnumbers,showpacs,superscriptaddress]{revtex4-2}
\usepackage{graphicx,color}
\usepackage{dcolumn}
\usepackage{bm}
\usepackage{braket}
\usepackage{mathtools}
\usepackage{ulem}
\usepackage[breaklinks,colorlinks=true,linkcolor=blue,urlcolor=blue,citecolor=blue]{hyperref}
\usepackage{times}
\usepackage{physics}
\usepackage{latexsym}
\usepackage{amsmath, amssymb}
\usepackage{mathtools}
\usepackage{multirow}

\newcommand{\be}{\begin{eqnarray}}
\newcommand{\ee}{\end{eqnarray}}

\renewcommand{\theequation}{\arabic{equation}}

\begin{document}
\title{R\'{e}nyi and Shannon mutual information in critical and decohered critical system}

\date{\today}
\author{Yoshihito Kuno} 
\affiliation{Graduate School of Engineering science, Akita University, Akita 010-8502, Japan}

\author{Takahiro Orito}
\affiliation{Department of Physics, College of Humanities and Sciences, Nihon University, Sakurajosui, Setagaya, Tokyo 156-8550, Japan}

\author{Ikuo Ichinose} 
\thanks{A professor emeritus}
\affiliation{Department of Applied Physics, Nagoya Institute of Technology, Nagoya, 466-8555, Japan}


\begin{abstract} 
We investigate a critical many-body system by introducing a R\'{e}nyi generalized mutual information, connecting between R\'{e}nyi mutual information and R\'{e}nyi Shannon mutual information.
This R\'{e}nyi generalized mutual information can offer more experimentally accessible alternative than the conventional entanglement entropy. 
As a critical many-body state, we focus on the critical transverse-field Ising model (TFIM) described by the Ising conformal field theory (CFT). 
We show that even if we modify the non-selective projective measurement assumed in R\'{e}nyi Shannon mutual information by replacing the measurement into decoherence by environment, 
the R\'{e}nyi generalized Shannon mutual information maintains the CFT properties such as subsystem CFT scaling law and its central charge observed through both the conventional R\'{e}nyi Shannon mutual information and R\'{e}nyi mutual information.
Furthermore, we apply a local decoherence to the critical ground state of the TFIM and numerically observe the R\'{e}nyi generalized mutual information 
by changing the parameter controlling environment effect (corresponding to the strength of measurement) in the R\'{e}nyi generalized mutual information and the strength of the decoherence to which the entire system subjects. 
We find that R\'{e}nyi-$2$ type central charge connected to the central charge is fairly robust, indicating the strong robustness of the Ising CFT properties against local decoherence 
by environment.
\end{abstract}


\maketitle
\section{Introduction}
Quantum information theoretic quantity characterize various non-trivial properties of quantum many-body states \cite{Wen_text}. Entanglement entropy is one of the most famous examples 
applied to various non-trivial quantum many-body states such as symmetry-protected topological states \cite{Ryu_Hatsugai}, many-body localization \cite{Bardarson2012} and topological ordered states \cite{Kitaev_Preskill,Levin_Wen}. 
Furthermore, critical many-body systems which can be described by conformal field theory (CFT)\cite{di1996conformal} is also characterized by various quantum information theoretic quantities. 
One of the most successful approaches is observing the subsystem dependence of entanglement entropy \cite{Calabrese_2004,Calabrese2009}. In particular,
by employing the entanglement entropies with different subsystem, the mutual information is defined. 
Then, the subsystem dependence of the mutual information shows a sine-function scaling form dubbed ``CFT scaling'' and its coefficient is a universal important factor namely ``central charge''. 
The central charge can be used to classify various critical ground states appeared on various many-body systems due to its universality. 
Even for some critical mixed states, such a sine-function scaling form can also be observed by using entanglement negativity \cite{Calabrese2012}. 

It is an important issue whether the critical properties of the system described by CFT with the distinctive central charge, 
are robust to various changes of the system such as adding interactions, applying disorders, and change of observational
protocol of physical quantities detecting the CFT properties. 
In many studies, the CFT properties has been examined by observing the bipartite entanglement entropy (EE) \cite{Calabrese_2004,Calabrese2009}. 
In particular, the cental charge deduced from the CFT view point is one of the most extensively studied quantities, and 
its numerical observation has been done for various models \cite{Vidal2003,Laflorencie2006,Tagliacozzo2008,Evenbly2009}.

Along this line, there are important studies \cite{Um_2012,Alcaraz2013,Alcaraz2014,Alcaraz2016} 
proposing and studying a Shannon mutual information (SMI), which is defined based on the Shannon entropy;
$S_S=-\sum_i p_i \log p_i$ with $p_i=|\langle i|\psi\rangle|^2$, where $|\psi\rangle$ is the ground state wave function and 
$\{|i\rangle\}$ is an arbitrarily-chosen complete basis.
The SMI exhibits a typical CFT scaling similar to the bipartite EE \cite{Calabrese_2004,Calabrese2009}. 
In most of critical ground states, the CFT scaling of the SMI can be given as,
$$
\frac{c}{4} \ln\left( \frac{L}{\pi} \sin\left( \frac{\ell \pi}{L} \right) \right) + \gamma,
$$
where $L$ is the system size (periodic boundary), $\ell$ is subsystem size, $c$ is a central charge and $\gamma$ is non-universal constant.
The R\'{e}nyi extension of the SMI, which is defined by replacing $S_S \to -\log (\sum_i p^n_i)$, was also proposed and it behaves similarly to R\'{e}nyi mutual information \cite{Alcaraz2014,Stephan2014,Alcaraz2015,Alcaraz2016}, that is,
$$
\frac{c_n}{4} \ln\left( \frac{L}{\pi} \sin\left( \frac{\ell \pi}{L} \right) \right) + \gamma,
$$
where $c_n$ is an extension of the central charge and depends on the R\'{e}nyi index $n$,
and the $n$-dependence of $c_n$ has been reported in \cite{Stephan2014,Alcaraz2014}. 
Hereafter, we call $c_n$ R\'{e}nyi-$n$ central charge.
In particular, for Ising CFT models, there exists a conjecture \cite{Alcaraz2014,Stephan2014,Alcaraz2015,Alcaraz2016}:
\begin{eqnarray}
c_n=
\left\{
\begin{array}{ll}
c, & (n=1) \\
c(\frac{n}{n-1}), & (n>1),\nonumber 
\end{array}
\right.
\end{eqnarray}
where the exact central charge of the Ising CFT, $c=1/2$ \cite{di1996conformal,Calabrese_2004}. This conjecture of $c_n$ has been numerically tested in detail\cite{Stephan2010,Alcaraz2013,Alcaraz2014,Stephan2014}. 
The numerical investigation of the SMI and its R\'{e}nyi ones \cite{Alcaraz2014,Stephan2014,Alcaraz2015,Alcaraz2016} indicates 
the robustness of CFT scaling and the central charge under the change of the physical condition applying the target system.
However, this robustness of the CFT properties is still a level of conjecture and there is no exact analytical proof even for some specific cases. 
Thus, further numerical investigation to clarify the regime in which the robustness holds is desired. 

As the first topic in this work, we propose an generalization of the R\'{e}nyi Shannon entropy. 
The formation of the Shannon type observables such as the Shannon entropy is based on specific projective measurements where we ourself determine the measurement bases \cite{Furukawa2009}. 
Here, we introduce the deformation, that is, relaxing the measurement strength in the Shannon type observables. 
Herein, we introduce a  R\'{e}nyi-2 generalized Shannon mutual information (R2GSMI). We investigate the R2GSMI for the critical ground state of the transverse field Ising model(TFIM) by employing an efficient numerical scheme to calculate the R2GSMI in detail. Interestingly enough, we find the fairly robustness of the CFT scaling and its central charge.

Moreover, for quantum systems, noise and decoherence from environment are inevitable~\cite{gardiner2000}. Investigation of effects of noise and decoherence have attracted significant attention in condensed matter and quantum information communities. In particular, how such effect changes quantum critical ground states, especially its criticality. 
As the second topic in this work, with the scope of the R2GSMI, we observe how the CFT scaling and its central charge behaves for a decohered critical mixed state created from the pure critical ground state of the TFIM by applying local decoherences. The study of the CFT behavior is attracted. Much recently, from other physical quantities the above issues has been investigated \cite{Zou2023,Ashida2024,KOI2025_1}. 
In this work, we numerically investigate 
the R2RGMI for the decohered mixed state made from the critical ground state docohered by $Y$-local decoherence channel. 
Under a fixed measurement bases (called conformal basis \cite{Alcaraz2014}) in the R2RGMI we numerically find a large regime of the Ising CFT in a system parameter space, where the Ising CFT scaling and its central charge are invariant.   

The rest of this paper is organized as follows. 
In Sec.~II, we start to show the target critical system, critical  TFIM in one dimension. 
In Sec.~III, physical quantities we focus on are introduced. We start to show R\'{e}nyi-2 Shannon entropy and we propose its generalized version, namely R2RSMI. We show the connection between other observables.
In Sec.~IV, we explain a practical numerical calculation way for the R2RSMI based on the doubled Hilbert space formalism, which can be efficiently tractable by employing matrix product states in a numerical calculation library \cite{TeNPy,Hauschild2024}. 
In Sec.~V, we show the systematic numerical investigation: 
(i) we observe the R2GSMI for the pure critical ground state of the TFIM and (ii) we observe the R2GSMI for the decohered critical ground state of the TFIM. 
Section VI is devoted to summery and conclusion.

\section{Target critical system}
In this work, our target model is 1D TFIM, 
represented by Pauli operators $X_j$, $Y_j$ and $Z_j$.
The target Hamiltonian is given by
\begin{eqnarray}
H_{\rm TFIM}=-\sum^{L-1}_{j=0}\biggr[Z_jZ_{j+1}+X_j\biggl],
\end{eqnarray}
the ground state of which is critical. 
Throughout this work, we consider periodic boundary conditions with a system size $L$. 
It is known that the critical ground states is described by Ising CFT \cite{di1996conformal}, that is, 
the critical ground state is expected to exhibit the conformal invariance and long-range properties of the TFIM model is described by fermion CFT, 
from which large distance exponent of correlation function is obtained analytically \cite{Giamarchi2003}.

\section{R\'{e}nyi generalized Shannon mutual information}
Throughout this work, we consider bi-partition of the system into two subsystems, $A$ and $B(=\bar{A})$, 
where the size of the subsystem $A$ is $L_A$ (including $L_A$ spins) 
and the size of the subsystem $B$ is $L_B=L-L_A$. 
In general, the pure many-body state $|\psi\rangle$ is expressed in terms of the two subsystem tensor basis,
\begin{eqnarray}
|\psi\rangle=\sum_{\ell_A,\ell_{B}}c_{\ell_A,\ell_B}|\ell_A\rangle_A|\ell_B\rangle_B,
\label{pure_state}
\end{eqnarray}
where $\{|\ell_{A(B)}\rangle\}$ is a set of orthogonal basis in the subsystem $A(B)$.

\subsection{R\'{e}nyi-2 Shannon entropy}
As the orthogonal states in Eq.~(\ref{pure_state}), we employ the diagonal states for a set of operators $\{\hat{M}_j\}$
such as $\{Z_j\}$ within the subsystem $A(B)$, that is, the eigenstates $\{|\ell^M_{A(B)}\rangle\}$ of the operator $\{\hat{M}_j\}$ within the subsystem $A(B)$.
Then,
by using the state representation of Eq.(\ref{pure_state}), R\'{e}nyi-2 Shannon entropy (R2SE) for the subsystem $A$ and the $M$-operator basis is expressed as \cite{Stephan2010,Stephan2011,Alcaraz2014} 
\begin{eqnarray}
SH^{(2)}_{A,M}=-\log\biggr[\sum_{\ell^M_A}(p^{A}_{\ell^M_A})^2\biggl],
\end{eqnarray} 
where $p^{A}_{\ell^M_A}$ is called $A$-subsystem marginal probability \cite{Alcaraz2013,Alcaraz2014} and 
given by $p^{A}_{\ell^M_A}\equiv \sum_{\ell^M_B}|c_{\ell^M_A,\ell^M_B}|^2$.

The R2SE can be regarded as an experimentally tractable observable compared to entanglement entropy at least for intermediate scale systems \cite{Alcaraz2013}, and 
the quantity $|c_{\ell^M_A,\ell^M_B}|^2$ is to be extracted by a specific projective measurement
with suitably chosen operator $\hat{M}$ \cite{Furukawa2009,Alcaraz2013}. 

\subsection{R\'{e}nyi-2 generalized Shannon entropy}

The R2SE depends on the choice of the orthogonal basis. 
It is related to the experimental manipulation of choosing a set of basis with which projective measurement is performed
on the subsystem region in order to obtain the target quantity, the marginal probability. 
From this measurement point of view, 
we generalize the R2SE by relaxing the constraint of such a projective measurement, that is, 
generalizing the perfect projective measurement to the one subject to noise, namely decoherence.
More comments will be given after explaining details of the proposal.

Based on this line, we propose the R\'{e}nyi-2 generalized Shannon entropy (R2GSE) for a subsystem $A$ \cite{Stephan2010,Alcaraz2014}. 
To this end, let us consider the density matrix of a pure state $\rho=|\psi\rangle\langle \psi|$, where $|\psi\rangle$ is a pure state. 
Then, we apply a local decoherence described by a local operator $\hat{M}$ to the subsystem $A$, 
\begin{eqnarray}
\mathcal{E}^{M,p_m}_{A}[\rho]&=&\prod_{j\in A}\mathcal{E}^{M,p_m}_{j}[\rho],\\
\mathcal{E}^{M,p_m}_{j}[\rho] &=& \biggr[(1-p_{m})\rho+p_{m}\hat{M}_{j}\rho \hat{M}_{j}\biggl], 
\end{eqnarray} 
where the strength of the decoherence is tuned by $p_{m}$. 
In the following we denote $\mathcal{E}^{M,p_m}_{A}[\rho]$ simply as $\rho^{M,A}_{p_m}$. 
Then, for the partially decohered state $\rho^{M,A}_{p_m}$, 
we take the partial trace for $\bar{A}$ and obtain the reduced density matrix,
\begin{eqnarray}
\rho^{M,A}_{A,p_m}=\Tr_{\bar{A}}[\rho^{M,A}_{p_m}].
\end{eqnarray} 
Then, for the state $\rho^{M,A}_{A,p_m}$, the R2GSE is defined as 
\begin{eqnarray}
S^{(2)}_{A,M}(p_m)\equiv -\log \Tr_{A}[(\rho^{M,A}_{A,p_m})^2].
\label{def_R2RSE}
\end{eqnarray}
Here, we note that 
for $p_m=0$, the R2GSE corresponds to the conventional R\'{e}nyi-2 entanglement entropy for the subsystem $A$, 
just obtained by the reduced density matrix $\rho_{0,A}=\Tr_{B}[\rho]$.
However, for $p_m\neq 0$ the R2GSE is different from the R\'{e}nyi-2 entanglement entropy since formally the R\'{e}nyi-2 entanglement entropy 
is calculated after we apply the partial decoherence of an operator $\hat{M}$ within the subsystem to the state $\rho$.
More detailed on this point is explained in the following subsection.

\subsection{Connection to R\'{e}nyi-2 Shannon entropy}
The R2GSE is reduced to the R\'{e}nyi-2 Shannon entropy in the limit $p_m\to 1/2$. 
By setting $p_m=1/2$, the decoherence is maximal and works as a non-selective projective measurement \cite{lidar2020}, and as a result, $\rho^{M,A}_{1/2}$ is expressed as,
\begin{eqnarray}
\rho^{M,A}_{1/2}=\sum_{\{|\ell^M_{A}\rangle\}}P^{M}_{|\ell^M_{A}\rangle}\rho P^{M}_{|\ell^M_{A}\rangle},
\end{eqnarray}
where $P^{M}_{|\ell^M_{A}\rangle_A}=|\ell^M_{A}\rangle_A {}_A\langle \ell^M_A|$. 

After some calculation, we obtain 
\begin{eqnarray}
\rho^{M,A}_{A,1/2}=\Tr_{B}[\rho^{M,A}_{1/2}]=\sum_{\ell^{M}_A}p^A_{\ell^{M}_A}|\ell^{M}_A\rangle_A {}_A\langle \ell^{M}_A|.
\end{eqnarray} 
Thus, by substituting $\rho^{M,A}_{A,1/2}$ into the R2GSE of Eq.~(\ref{def_R2RSE}), we obtain
\begin{eqnarray}
S^{(2)}_{A,M}(1/2)&=&-\log \Tr[(\rho^{M,A}_{A,1/2})^2]=-\log\biggr[\sum_{\ell^M_A}(p^{A}_{\ell^M_A})^2\biggl]\nonumber\\
&=&SH^{(2)}_{A,M}.
\end{eqnarray}

From the above argument, one may wonder what is physical meaning of the parameter $p_m$ from the view point of Shannon entropy.
As we explained in the above, Shannon entropy is obtained by observing the target state with a set of local operators $\{\hat{M}_j\}$.
Through the perfect measurement, the exact value of Shannon entropy of the state is obtained, whereas in general, measurement is noisy,
and obtained amplitudes, $p$'s, contain error.
In our formalism of the R2GSE, the parameter $p_m$ refers to magnitude of noise and accuracy of the measurement.
We think that this point of view becomes important for practical measurement of Shannon entropy in experiment on intermediate scale device.
In fact, the numerical study shown later on will provide a benchmark for tolerance toward noise from 
CFT aspect of Shannon entropy.

\begin{figure}[t]
\begin{center} 
\vspace{0.5cm}
\includegraphics[width=6.5cm]{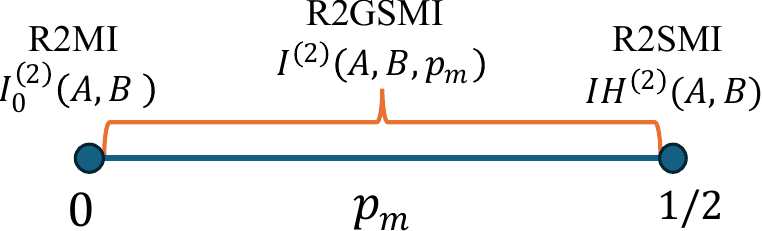}  
\end{center} 
\caption{Relation between the mutual information considered in this work. 
The R2GSMI with the parameter $p_m$ can be regarded as the intermediate mutual information connecting the R2MI ($p_m=0$) and R2SMI ($p_m=1/2)$.}
\label{Fig0}
\end{figure}
\subsection{Mutual information}
We next define two types of mutual information. 
The first one is the R\'{e}nyi-2 Shannon mutual information (R2SMI) introduced in \cite{Alcaraz2014,Stephan2014},
\begin{eqnarray}
IH^{(2)}(A,B)&\equiv& SH^{(2)}_{A,M}+SH^{(2)}_{B,M}-SH^{(2)}_{A\cup B,M}.
\label{R2SMI}
\end{eqnarray}

The second one is R\'{e}nyi-2 generalized Shannon mutual information (R2GSMI), which is an extension of the R2SMI and we originally introduce it in this work, 
\begin{eqnarray}
I^{(2)}(A,B,p_m)\equiv S^{(2)}_{A,M}(p_m)+S^{(2)}_{B,M}(p_m)-S^{(2)}_{A\cup B,M}(p_m).\nonumber\\
\label{R2RSMI}
\end{eqnarray}
Here we comment that $S^{(2)}_{A\cup B,M}(p_m)$ is nothing but system environment entanglement \cite{Ashida2024}, and
also, 
$$
\displaystyle{\lim_{p_{m}\to 1/2}I^{(2)}(A,B,p_m)=IH^{(2)}(A,B)}.
$$
Then, in the limit, $p_m\to 0$, 
the R2GSMI reduces to the R\'{e}nyi-2 mutual information (R2MI) denoted by $I^{(2)}_0(A,B)$,
$$
\lim_{p_{m}\to 0}I^{(2)}(A,B,p_m)= I^{(2)}_0(A,B).
$$
From these two sides of the limits, the R2GSMI is an intermediate physical quantity between the R2MI and the R2GSMI. 
We show the image of the relation among three types of mutual information considered in this work in Fig.\ref{Fig0}.

In this work, our main focus is the R2GSMI, $I^{(2)}(A,B,p_m)$. 
We show numerical methods for calculation of the R2GSMI in the following section.
We also emphasis that even if the state $\rho$ is a decohered mixed state, the R2GSMI can be calculated without any difficulties. 

\section{Practical numerical calculation scheme of R2GSMI}

In this section, we show an efficient numerical scheme to calculate the R2GSE for a density matrix by using the doubled Hilbert space formalism. 

In general, mixed state density matrices (including pure states) $\rho\in \mathcal{H}$, where $\mathcal{H}$ is a target Hilbert space, 
are treated like a pure state by the doubled Hilbert space formalism \cite{Lee2025}. 
The target Hilbert space $\mathcal{H}$ is doubled as $\mathcal{H}_{u}\otimes \mathcal{H}_{\ell}$, 
where the subscripts $u$ and $\ell$ refer to the upper and lower Hilbert spaces corresponding to the ket and bra states of mixed state density matrix, respectively. 
Then, the 1D system we consider can be regarded as a ladder system. 
The density matrix $\rho$ becomes a ``supervector'' as 
$\rho \longrightarrow |\rho\rangle\rangle\equiv \frac{1}{\sqrt{\dim[\rho]}}\sum_{k}|k\rangle\otimes \rho|k\rangle$, where $\{|k\rangle \}$ 
is an orthonormal set of basis in the Hilbert space $\mathcal{H}$. This is just the Choi-Jamio\l kowski isomorphism \cite{Choi1975,JAMIOLKOWSKI1972}. 

We show how to calculate the R2GSE, $S^{(2)}_{A,M}(p_m)$ from the supervector $|\rho\rangle\rangle$.
We start with applying the partial decoherence $\mathcal{E}^{M,p_m}_{A}$ to the state $\rho$. 
This can be easily carried out in the doubled system by replacing the channel $\mathcal{E}^{M,p_m}_{A}$ 
into the operators acting on the state $|\rho\rangle\rangle$ (See Appendix) $\mathcal{E}^{M,p_m}_{A}\longrightarrow \hat{\mathcal{E}}^{M,p_m}_{A}$. 
Then, $\rho^{M,A}_{p_m}$ is obtained as a supervector $|\rho^{M,A}_{p_m}\rangle\rangle \equiv \hat{\mathcal{E}}^{M,p_m}_{A}|\rho\rangle\rangle$.

Second, we need to calculate the reduced density matrix $\rho^{M,A}_{A,p_m}=\Tr_{B}[\rho^{M,A}_{p_m}]$ and the trace of its squared quantity,
$\mbox{Tr}_{B}[(\rho^{M,A}_{A,p_m})^2]$. 
These steps can be efficiently carried out \cite{KOI2025_5}. 
We further divide each Hilbert space into the subsystems $A$ and $B$ as $\mathcal{H}_{u(\ell)}=\mathcal{H}_{A,u(\ell)}\otimes \mathcal{H}_{B,u(\ell)}$. 
Tracing out the degrees of the freedom in the subsystem $B$ and taking trace of the squared reduced density matrix $\Tr_{A}[(\rho^{M,A}_{A,p_m})^2]$ 
are carried out in the following way;
For the supervector $|\rho^{M,A}_{p_m}\rangle\rangle$, we apply a suitable maximal depolarization channel to the subsystem $B$ \cite{zhang2025_SP}. 
The maximal depolarization channel \cite{Nielsen2011} is represented by an operator (the same as $\hat{\mathcal{E}}^{M,p_m}_{A}$) 
acting on the vector $|\rho^{M,A}_{p_m}\rangle\rangle$ \cite{Lee2025}, which is given by
\begin{eqnarray}
\hat{D}_{B}&=&\prod_{j\in B}\hat{D}^{m}_j,\\
\hat{D}^{m}_j&=&\frac{1}{4}\biggr[\hat{I}_{j,u} \otimes \hat{I}_{j,\ell}+\hat{X}_{j,u} \otimes \hat{X}_{j,\ell}-\hat{Y}_{j,u}\otimes \hat{Y}_{j,\ell}\nonumber\\
&&+\hat{Z}_{j,u} \otimes \hat{Z}_{j,\ell}\biggr],
\end{eqnarray}
where $\hat{I}_{j,u(\ell)}$ is an identity operator for site-$j$ vector space in $\mathcal{H}_{u(\ell)}$, $Z(X,Y)_{j,u(\ell)}$ is Pauli-$Z$($X$,$Y$) operator at site $j$. 
The depolarization operator $\hat{D}_{B}$ is regarded as a coupling between the system $u$ and $\ell$ \cite{KOI2025_5}. 
Then, we act $\hat{D}_{B}$ to $|\rho^{M,A}_{p_m}\rangle\rangle$ and obtain the following identity~\cite{zhang2025_SP}, 
\begin{eqnarray}
\hat{D}_{B}|\rho^{M,A}_{p_m}\rangle\rangle=|\frac{{\bf I}_{B}}{d_{B}}\otimes \rho^{M,A}_{A,p_m}\rangle\rangle,
\label{MD_1}
\end{eqnarray}
where $d_{B}$ is the number of the degree of freedom of the subsystem $B$ ($d_{B}=2^{L-L_A}$). 
Then, there is correspondence such as
$|\frac{{\bf I}_{B}}{d_{B}}\otimes \rho^{M,A}_{A,p_m}\rangle\rangle \longleftrightarrow \frac{{\bf I}_{B}}{d_{B}}\otimes [{\rm Tr}_{B}[\rho^{M,A}_{p_m}]]$, 
meaning that the maximal partial depolarization projects the state of the subsystem $B$ onto the infinite temperature state, 
that is, all information in the subsystem $B$ is swept out.

As the final step, the inner product of the doubled Hilbert space vectors $\langle\langle A|B\rangle\rangle$ corresponds to 
$\Tr[A^\dagger B]$ \cite{Ma2024_double}. 
Then, through the norm of $\hat{D}_{B}|\rho^{M,A}_{p_m}\rangle\rangle$, 

$\displaystyle{
\langle\langle \frac{{\bf I}_{B}}{d_{B}}\otimes \rho^{M,A}_{A,p_m} |\frac{{\bf I}_{B}}{d_{B}}\otimes \rho^{M,A}_{A,p_m}\rangle\rangle 
=\frac{1}{d_{B}}{\rm Tr}_{A}[(\rho^{M,A}_{A,p_m})^2]
}
$.
Then, we obtain the R2GSE $S^{(2)}_{A,M}(p_m)$ as
\begin{eqnarray}
S^{(2)}_{A,M}(p_m)=-\log\biggr[d_{B}\langle\langle \frac{{\bf I}_{B}}{d_{B}}\otimes \rho^{M,A}_{A,p_m} |\frac{{\bf I}_{B}}{d_{B}}\otimes \rho^{M,A}_{A,p_m}\rangle\rangle \biggl].\nonumber\\
\end{eqnarray}

These numerical calculations above can be easily performed by the filtering methods of the matrix product state presented in previous studies \cite{Orito2025,kuno_2025,kuno_2025_v1}. Thus, by changing the choice of the subsystem, we can numerically obtain R2GSEs with different subsystem size, $S^{(2)}_{A,M}(p_m)$, $S^{(2)}_{B,M}(p_m)$ 
and $S^{(2)}_{A \cup B,M}(p_m)$. As a results, we obtain the R2GSMI, $I^{(2)}(A,B,p_m)$.

\section{Numerical results}
In this section, we show the numerical results of the R2GSMI for two cases.
We used the calculation scheme explained in the previous section and employed the DMRG \cite{TeNPy,Hauschild2024} 
to prepare the critical ground state of the TFIM.

For some typical cases, we examine the CFT scaling law from the numerical data of the R2GSMI, the ansatz of which is given by 
\begin{eqnarray}
I^{(2)}(A,B,p_m)=\frac{c_2}{4} \ln\left( \frac{L}{\pi} \sin\left( \frac{L_A \pi}{L} \right) \right)+b_2.
\label{anzats}
\end{eqnarray}
We extract the estimated values $c_2$ and $b_2$. 
In what follows, we call $c_2$ R\'{e}nyi-$2$ central charge as the specific case $n=2$.
By the conjecture \cite{Alcaraz2014,Stephan2014,Alcaraz2015,Alcaraz2016}, the central charge $c$ is related to the R\'{e}nyi-$2$ central charge as $c_2=2c$.

Although for the critical ground state of the TFIM, the previous works numerically verified $c_2=1$($c=1/2$) in the limit $p_m\to 1/2$ \cite{Alcaraz2014,Alcaraz2016,Stephan2014}, that is, the exact value of the Ising CFT, $c=1/2$ \cite{di1996conformal,Calabrese_2004} was observed. 
However, the global behavior $c_2$ estimated from the R2GSMI is lacking. 

As technical aspects in this section, we set $L=32$ and use the numerical data of $I^{(2)}(A,B,p_m)$ with $8\leq L_A\leq 24$. 
We prepare the double chain of the critical ground state of the TFIM by using the DMRG algorithm of the TeNPy library \cite{TeNPy,Hauschild2024} for the calculation of the $I^{(2)}(A,B,p_m)$. 
In this DMRG calculation, for preparing the initial critical ground state, we set maximum bond dimension $D=200$-$240$, truncate the singular value less 
than $\mathcal{O}(10^{-7})$, and  keep the energy convergence of the iterative DMRG sweep as $\Delta E < \mathcal{O}(10^{-6})$ to obtain the initial MPS ground state. 
We employ these numerical conditions in the rest of the present work.  

\begin{figure}[t]
\begin{center} 
\vspace{0.5cm}
\includegraphics[width=9cm]{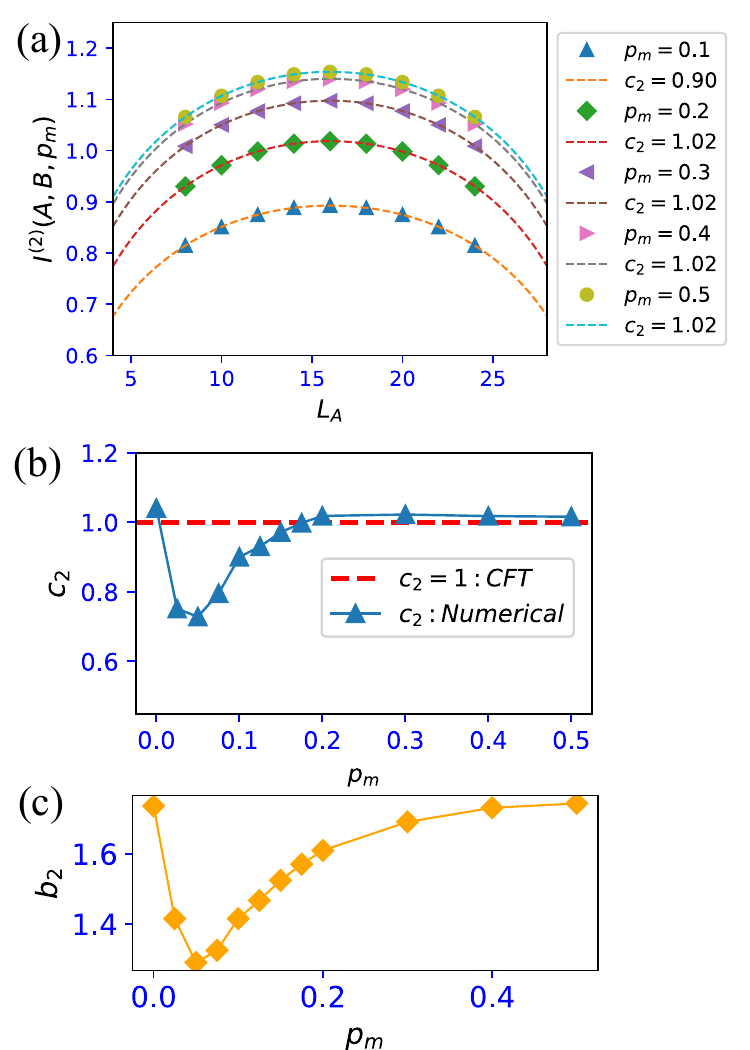}  
\end{center} 
\caption{(a) $L_A$-dependence of the R2GSMI for $\hat{M}=Z$ conformal basis. (b) $p_m$-dependence of estimated central charges. 
(c) $p_m$-dependence of estimated non-universal offsets in CFT scaling form}
\label{Fig1}
\end{figure}

\subsection{Case I: R2GSMI for pure critical ground state}
Let us numerically investigate how the CFT scaling and R\'{e}nyi-$2$ central charge, $c_2$, are preserved or changed in relaxing the projective measurement in the R2SMI. 
That is, we observe $p_m$-dependence of R2GSMI.

\underline{$\hat{M}=Z$ bases}: 
We start with studying the R2GSMI of the system subject to the $Z$-basis measurement, $\hat{M}=Z$, which corresponds to the conformal boundary of the Ising CFT.
The TFIM is connected to the two-dimensional (2D) Ising model, and $Z$-basis projection in the TFIM is equivalent to fixing the spins in the classical counterpart.
As discussed in \cite{Stephan_2014_boundary}, this boundary condition is a conformal boundary condition of the CFT.

\begin{figure}[t]
\begin{center} 
\vspace{0.5cm}
\includegraphics[width=9cm]{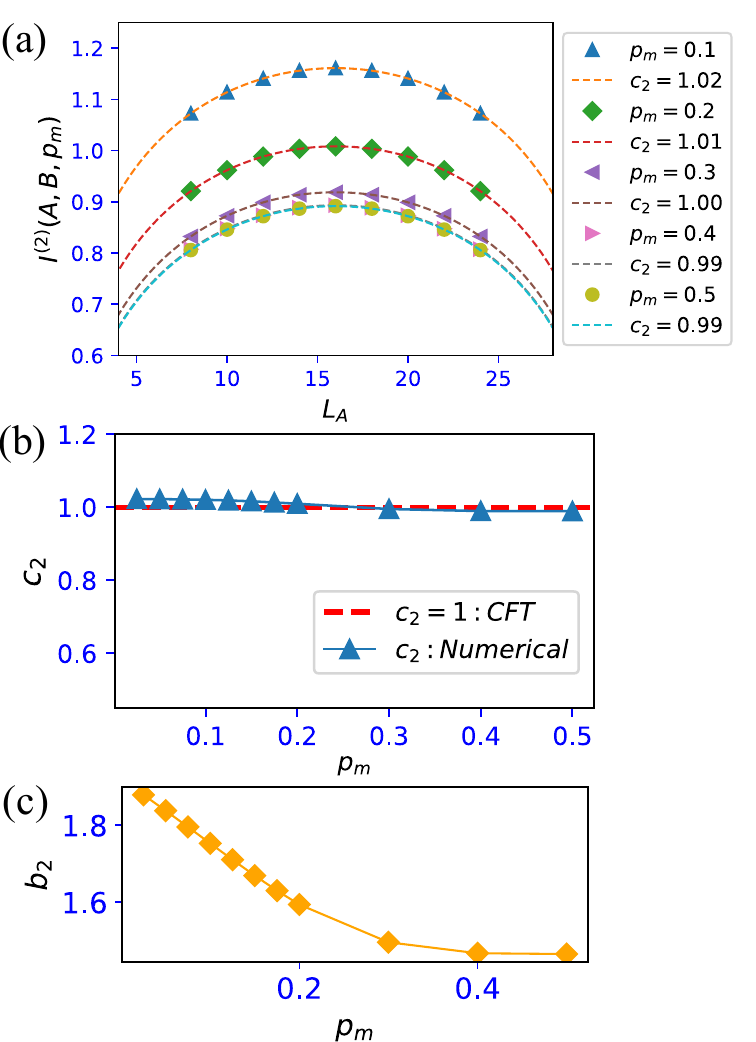}  
\end{center} 
\caption{(a) $L_A$-dependence of the R2GSMI. (b) $p_m$-dependence of estimated central charges. 
(c) $p_m$-dependence of estimated non-universal offsets in CFT scaling form. For all data, for $\hat{M}=X$ conformal basis is used.}
\label{Fig2}
\end{figure}

Figure \ref{Fig1}(a) shows the scaling data for typical $p_m$'s. 
We find that all cases are well-fitted with the ansatz of Eq.~(\ref{anzats}). 
That is, the CFT scaling form is robust against the relaxation by $p_{m}$. 
However, we find that the estimated R\'{e}nyi-$2$ central charge $c_2$ exhibits an interesting $p_m$-dependence as shown in Fig.~\ref{Fig1} (b). 
In $0.2 \leq p_m \leq 0,5$, the value of $c_2$ keeps the value of the maximal-projection case conjectured in \cite{Alcaraz2014,Stephan2014}, i.e., $c_2=1$. 
This means that the R\'{e}nyi-$2$ central charge $c_2=1$ obtained by the numerical work of the R2SMI in \cite{Alcaraz2014} 
is robust against the relaxation of the projective measurement up to $p_m=0.2$. 
On the other hand, interestingly enough for $0< p_m <0.2$, the R\'{e}nyi-$2$ central charges deviate from $c_2=1$ continuously, that is,
the R2RSMI {\it loses} the connection with Ising CFT central charge or the conjecture $c_2=1$($c=1/2$). 
However, the point $p_m=0$ is typical. 
The R2RSMI reduces R\'{e}nyi-2 mutual information, which is $I^{(2)}(A,B,p_m)=2S^{(2)}_{A,Z}(0)$ since the critical ground state is translational invariant and $S^{(2)}_{A\cup B,Z}(0)=0$. 
At this point, an exact result of the Ising CFT \cite{Calabrese_2004} is known, $c_2=1$ ($c=1/2$). 
As our numerical results show, the value of $c_2$ approaches unity again, exhibiting the clear connection to the exact value of the central charge of the Ising CFT \cite{Calabrese_2004}. 

Moreover, as a byproduct, we also observe the non-universal term $b_2$ estimated by the scaling ansatz. 
The result is shown in Fig.~\ref{Fig1} (c). 
The behavior of $b_2$ is similar to that of $c_2$. 
Around $p_m=0.1$, the value takes minimum and exhibits asymptotic behavior to the both sides of the limits $p_m\longrightarrow 0$ and $0.5$. 
But for $p_m > 0.2$, there is no plateau value compared to $c_2$.

\underline{$\hat{M}=X$ bases}:
We next move to the projection by the product $X$-basis, $\hat{M}=X$, which is known as the conformal boundary.
In fact, this $X$-basis projection corresponds to the free boundary condition in the 2D classical Ising model,
which is a conformal boundary of the CFT as discussed in \cite{CARDY1989_581}. 

Figure \ref{Fig2}(a) shows the scaling data for typical $p_m$'s. All data are well-fitted with the ansatz of Eq.~(\ref{anzats}). 
The CFT scaling form is robust against the relaxation by $p_{m}$. 
Furthermore, compared to the $Z$-basis, the estimated R\'{e}nyi-$2$ central charge $c_2$ exhibits no $p_m$-dependence, keeping $c_2=1$($c=1/2$) as shown in Fig.~\ref{Fig2} (b). 
This seems indicate that the critical properties of the TFIM described by the CFT remain unaffected up to the point of maximal decoherence (corresponding to the R2SMI) 
without any qualitative changes of the R2RSMI in $X$-basis. 
We expect that this behavior comes from the fact that $X$-basis measurement corresponds to the free boundary condition from the view point of the CFT as mentioned in the above.

\subsection{Case II: R2RSMI for decohered critical ground state}

We investigate decohered critical ground states (mixed state) in this subsection. 
We set $\hat{M}=Z$ in the calculation of the R2GSMI. In particular, we focus on the effects of a decoherence to the critical ground state and study the R2GSMI.
As the previous work \cite{Zou2023} shows, the TFIM critical ground state subject to local decoherence applied to the whole system exhibits the same scaling form to Eq.~(\ref{anzats}) for the R2MI. 
We numerically investigate how such a tendency is related to the R2GSMI.

In the following, we consider local $Y$-decoherence applied to the whole system, the operator type of which is different from the operator $\hat{M}(=Z)$ 
employed in the partial decoherence to observe the R2RSMI. 
For the decohered state, it is interesting to observe the behavior of the R2RSMI in the $(p_m, p_y)$ parameter plane. 

The local $Y$-decoherence applied to every site of the critical ground state of the TFIM $\rho=|\phi_0\rangle\langle \phi_0|$, the quantum channel of which is given by 
\begin{eqnarray}
\mathcal{E}^{Y}_{tot}[\rho]&=&\biggl(\prod^{L-1}_{j=0}\mathcal{E}^{Y}_{j}\biggl)[\rho]\equiv \rho_D,\\
\mathcal{E}^{Y}_{j}[\rho]&\equiv& (1-p_{y})\rho+p_{y}Y_{j}\rho Y_{j},
\end{eqnarray}
where the strength of the decoherence is tuned by $p_{y}$ ($j$-independent), and $0\leq p_{y}\leq 1/2$. 
How the R2RSMI for the mixed state $\rho_D$ behaves is a target of the present study.
For $p_{y}= 1/2$, the channel corresponds to the non-selective projective measurement of $Y_j$, which is called maximal decoherence.
This decohered state is also constructed as a supervector in the doubled Hilbert space formalism. 
Then, we just numerically calculate the R2RSMI for the supervector. 
Indeed, we have the two parameters $p_m$ and $p_y$ and investigate the R2RSMI.

\begin{figure}[t]
\begin{center} 
\vspace{0.5cm}
\includegraphics[width=8cm]{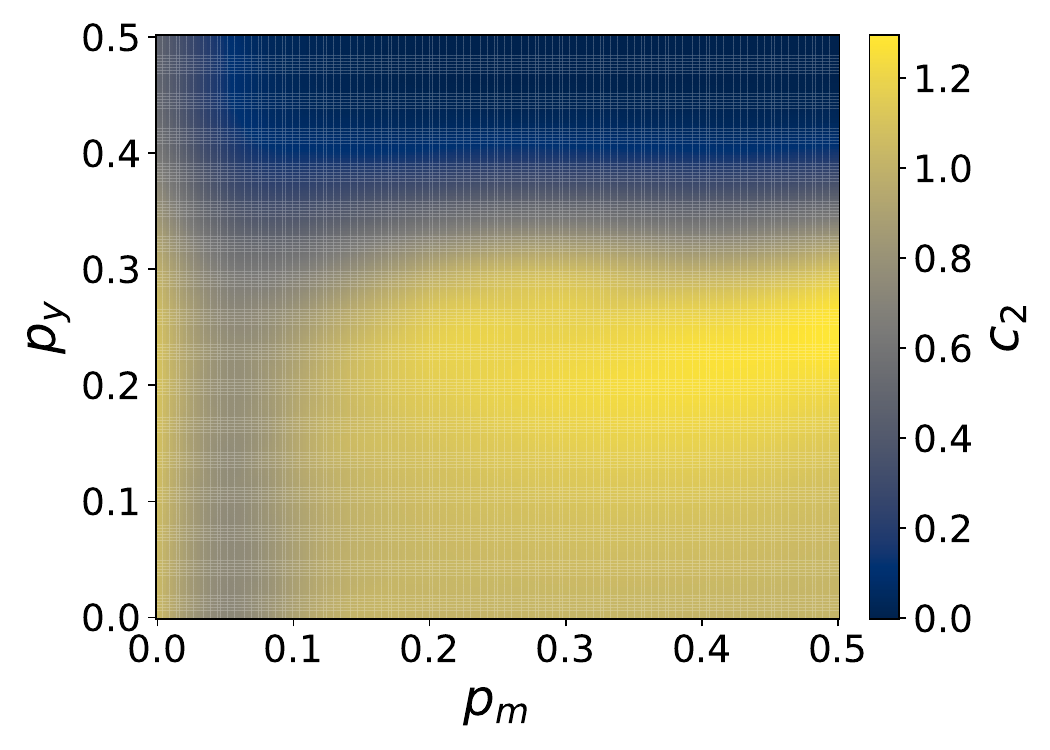}  
\end{center} 
\caption{Numerical extracted R\'{e}nyi-2 central charge $c_2$ from R2RSMI in $p_m$-$p_y$ plane. On $p_m=0$ line, R2RSMI corresponds to the R2RMI for decohered critical ground states. 
In this colormap, we used a cubic interpolation from the discrete data points.
}
\label{Fig3}
\end{figure}

Figure \ref{Fig3} shows the summarized results of the R\'{e}nyi-2 central charge $c_2$ numerically extracted by the ansatz of Eq.~(\ref{anzats}), 
where in the fitting of the ansatz of Eq.~(\ref{anzats}) we employed $L_A=8,12,14,16$ to obtain the sufficient accurate results.
We here show the global ($p_m$,$p_y$)-dependence of $c_2$. We find that even for both relaxation and $Y$-decoherence, 
the extracted R\'{e}nyi-$2$ central charge $c_2$ exhibits near Ising CFT value $c_2\sim 1$ shown as a broad regime in the $p_m$-$p_y$ plane in Fig.~\ref{Fig3}. 
This also indicates that the R2RSMI for both relaxation and $Y$-decoherence exhibits CFT scaling form and its Ising CFT central charge. 
Precisely, around $p_m=0.05$ line, at $p_{y}=0$, $c_2$ decreases in the R2GSMI of the $Z$-basis. as shown in Fig.\ref{Fig1}(b) 
and this tendency continuously keeps in a finite $p_y$. Also, for any relaxation (any $p_m$ fixed line), the strong $Y$-decoherence (large $p_y$) induces a deviation from the exact Ising CFT $c_2=1$ ($c=1/2$). 
In particular, in the non-selective projective  measurement limit $p_y=1/2$, its deviation tendency gets large for larger $p_m$.

In addition, at the parameter point with $p_m=0$ and $p_y=0$, the extracted R\'{e}nyi-$2$ central charge $c_2$ is tiny larger than exact one. 
This agrees to the previous numerical result in the previous work \cite{Stephan2014}. 
There, $c_2$ was numerically estimated by considering a free fermion CFT model(XY model) and for $Z$-basis, $c_2$ was estimated by $c_2=1.02$.

We also comment that the case $p_m=0$ of our numerics agrees with the numerical results in the previous work \cite{Zou2023}. 
The R2RSMI for the case $p_m=0$ with a finite $p_{y}$ has been already studied since the R2RSMI corresponds to the R\'{e}nyi-2 mutual information for the $Y$-decohered state with the decoherence strength $p_y$. 
We observe that our calculation in $p_m=0$ line exhibits fairly similar behavior to the one reported in \cite{Zou2023}, that is the R\'{e}nyi-2 central charge $c_2$ keeps $\sim 1$ up to a decoherence strength $p_y\sim 0.3$ 
and then the value of $c_2$ decays, indicating that the Ising CFT is robust up to the decoherence $p_y\sim 0.3$.

We summarize the numerical results in this section. 
We found that the CFT scaling and the central charge of Ising CFT have broad robustness to the change of the projective measurement in the R2SMI and local decoherences. 
As shown in the previous study \cite{Stephan2014}, this analytical derivation is difficult since this model is a fermion CFT, while in the bosonic CFT models, 
the appearance of the CFT scaling in the R2MI is analytically verified by a field theory.

\section{Conclusion}
In this work, we proposed a generalized mutual information, namely R2GSMI. 
This R2GSMI can be regarded as a relaxed version of the projective measurement assumed in the R2SMI. 
Indeed, we elongate the projective measurement assumption into the corresponding decoherence with a tuning parameter $p_m$. Thus, since the R2SMI can be an experimentally accessible observable as mentioned in \cite{Alcaraz2013}, 
the R2GSMI can have beneficial points from experimental points of view. 
In addition, the R2GSMI connects to the R2MI which is also a typical entanglement measure well-analyzed in the conformal field theory \cite{Calabrese_2004}. 
In this sense, the R2GSMI we proposed is an intermediate observable connecting the R2SMI and R2MI regimes. 

We numerically investigated the R2GSMI for the critical ground state of the TFIM. 
We found that the CFT scaling appears in a broad parameter regime $p_m$ for some basic measurement basis (conformal basis) and our estimated 
The R\'{e}nyi-2 central charges are fairly close to the exact central charge of the Ising CFT \cite{di1996conformal,Calabrese_2004}. 
So far, such a robustness for the R2SMI has been numerically investigated, however, our numerics showed that a further generalized mutual information exhibits such a robustness of the Ising CFT. 

We then investigate the R2GSMI for a decohered mixed critical ground state of the TFIM. 
There, we considered an entire $Y$-local decoherence channel. 
While so far the critical CFT behavior has been investigated in Ref.\cite{Zou2023} through the R2MI, we further investigated the CFT properties of the critical mixed state with the scope of the R2GSMI. 
In particular, through numerically extracting the R\'{e}nyi-2 central charge from the R2GSMI, we found that the Ising CFT tendency appears even in a broad finite decohered regime, 
indicating that the central charge of Ising CFT expected in the original CFT \cite{di1996conformal} is significantly invariant to the change of the mutual information measure and local decoherences.

In this work, we focused on the fermionic CFT. Investigation of the R2GSMI for bosonic CFT models such as XXZ chain is a significant future problem.\\

\section*{Acknowledgements}
This work is supported by JSPS KAKENHI: JP23K13026(Y.K.) and JP23KJ0360(T.O.). 

\section*{Data availability}
The data that support the findings of this study are available from the authors upon reasonable request.\\

\renewcommand{\thesection}{A\arabic{section}} 
\renewcommand{\theequation}{A\arabic{equation}}
\renewcommand{\thefigure}{A\arabic{figure}}
\setcounter{equation}{0}
\setcounter{figure}{0}

\appendix
\section*{Appendix}

\subsection*{Choi representation for decoherence channels}
In general, quantum decoherences are expressed in terms of Kraus operator form through Stein-spring representation. 
For a density matrix, the description is ~\cite{Nielsen2011,lidar2020}$$\mathcal{E}[\rho]=\sum^{M-1}_{\alpha=0}K_{\alpha}\rho K^\dagger_{\alpha},
$$ 
where $K_\alpha$'s are Kraus operators satisfying $\sum^{M-1}_{\alpha=0}K_{\alpha}K^{\dagger}_{\alpha}=I$.
In the Choi isomorphism \cite{Choi1975,JAMIOLKOWSKI1972}, 
the channel is also transformed as ~\cite{Lee2025} 
$$
\mathcal{E}\longrightarrow \hat{\mathcal{E}}=\sum^{M-1}_{\alpha=0}K^*_{\alpha,u}\otimes K_{\alpha,\ell}.$$
That is, the quantum channel represented by the Kraus operators becomes generally a non-unitary operator for the mixed state supervector in the doubled Hilbert space. 
Since the channel operator $\hat{\mathcal{E}}$ is not a unitary map in general cases although the channel is a completely positive trace-preserving map, 
the application of the channel operator generally changes the norm of the mixed state supervector. 
In fact, the norm is nothing but the purity of the original mixed state \cite{Ashida2024}.


\bibliography{ref}

\end{document}